\documentclass[twocolumn,showkeys,showpacs]{revtex4}
\usepackage{amsmath}
\usepackage{amsfonts}
\usepackage{amssymb}
\usepackage{graphicx}
\usepackage{pdfpages}

\begin{document}

\title{
Novel Effect Induced by Spacetime Curvature  
in Quantum Hydrodynamics 
}
\author{T.\ Koide}
\email{tomoikoide@gmail.com,koide@if.ufrj.br}
\affiliation{Instituto de F\'{\i}sica, Universidade Federal do Rio de Janeiro, C.P. 68528,
21941-972, Rio de Janeiro, RJ, Brazil}
\author{T.\ Kodama}
\email{kodama.takeshi@gmail.com,tkodama@if.ufrj.br}
\affiliation{Instituto de F\'{\i}sica, Universidade Federal do Rio de Janeiro, C.P. 68528,
21941-972, Rio de Janeiro, RJ, Brazil}
\affiliation{Instituto de F\'{\i}sica, Universidade Federal Fluminense, 24210-346,
Niter\'{o}i, RJ, Brazil}

\begin{abstract}
The interplay between quantum fluctuation and spacetime curvature is shown to induce an additional 
quantum-curvature (QC) term in the energy-momentum tensor of fluid 
using the generalized framework of the stochastic variational method (SVM).
The QC term is necessary to satisfy the momentum conservation but the corresponding 
quantum hydrodynamics is not necessarily cast into the form of the Schr\"{o}dinger equation, differently from the case of the Euclidean spacetime. 
This  seems to suggest that the existence of the Hilbert space 
is not a priori requirement in the quantization of curved spacetime systems.
As an example, we apply the Friedmann-Robertson-Walker (FRW) metric and  
show that this term contributes to the cosmological acceleration although it is too small in the present non-relativistic toy model.
\end{abstract}

\keywords{quantization, curved spacetime, variational principle, stochastic calculus}

\pacs{02.50.Ey,03.65.Ca,11.10.Ef,98.80.Qc}

\maketitle

\section{introduction}

There exists no established formulation of quantum mechanics in the curved spacetime.
Normally the formulation in the Euclidean spacetime is assumed to be held without a major modification. 
For example, the existence of the Hilbert space is required from the beginning. 
This is fairly natural but not trivial a priori.

The stochastic variational method (SVM) is one of quantization schemes \cite{yasue-81,zam-review,koide-review}. 
Quantization is then formulated as the stochastic optimization of classical actions: 
the Schr\"{o}dinger equation is derived by applying the stochastic variation to the action 
which leads to the Newton equation under the standard classical variation.
As a remarkable feature, the existence of the Hilbert space is not necessarily required. 
Instead, the optimized dynamics is represented in the Madelung-Bohm-type hydrodynamic form 
(quantum hydrodynamics) \cite{book:holl}, 
and it can be cast into the Schr\"{o}dinger equation only when the fluid velocity is expressed by the gradient of a velocity potential. 
If such a hydrodynamics has an additional term, 
the velocity potential is not introduced in general 
and then the quantized dynamics is not expressed with the wave function 
(See below Eq.\ (\ref{eqn:vari-particle}) and Appendix \ref{app:1}). 
Thus it is interesting to study the application of SVM in the curved spacetime to investigate 
the existence of the Hilbert space formed by the wave function.

The purpose of this paper is to develop the generalized SVM applicable to curved spacetime systems, assuming that the variational principle is a fundamental requirement 
in quantization.  
For the sake of simplicity, we consider non-relativistic systems of particle and continuum medium \cite{chavanis}.
Then we show that the interplay between quantum fluctuation and spacetime curvature induces 
the quantum-curvature (QC) term  
which can prevent us from introducing the wave function but is necessary to find a conserved energy-momentum tensor.
As an example of curved spacetime systems, we apply the Friedmann-Robertson-Walker (FRW) metric and discuss 
that the effect of the QC term contributes to the cosmological acceleration.

In this paper, $\hbar$, $c$ and $G$ denote the Planck constant, the speed of light and the gravitational constant, respectively. 
The stochastic quantity is denoted by $(~\hat{~}~)$.
The difference $dA(t)$ is $A(t+dt) - A(t)$, independently of the sign of $dt$.
The Einstein notation of the summation is used.

\section{formulation}

For a curved spacetime characterized by the metric $g_{\mu\nu}$, it is possible to find a local Minkowskian system with the metric $\eta_{ab}$ 
($= {\rm diag}( -1,1,1,1 )$). 
The spacetime position in the former general coordinate is denoted by $x^{\mu}$ while that in the latter by $y^{a}$. 
The Greek indices $\alpha,\beta,\cdots$ are used to label the general coordinate.
The Latin indices $a,b,\cdots$ are for the local Minkowskian coordinate but $i,j,k$ are reserved to denote the spatial components of $x^\mu$.
Then the tetrads \cite{synge} are  
$\underline{e}^\mu_a (X) = \left. \partial x^\mu/\partial y^{a} \right|_{x = X}$ and 
$\overline{e}^{a}_\mu (X) = \left. \partial y^a/\partial x^{\mu} \right|_{x = X}$, satisfying 
$g_{\mu\nu}\underline{e}^{\mu}_a \underline{e}^{\nu}_b = \eta_{ab}$ and $\eta_{ab}\overline{e}^{a}_\mu \overline{e}^{b}_\nu = g_{\mu\nu}$.
Note that $g = {\rm det}(g_{\mu\nu})$.

In the SVM  quantization, quantum effects are introduced through stochastic motions. 
We consider a non-relativistic Brownian motion in the curved spacetime where 
the time component is given by $dx^0_t= cdt$, and a limited spacetime geometry satisfying 
\begin{eqnarray}
g_{0i} = 0\, , \ \ \partial_i g_{00} = 0\, . \label{eqn:cond-geo}
\end{eqnarray}
The stochastic differential equation (SDE) of such a Brownian motion is already known for the curved space \cite{zam-review,ito}. 
The same equation is applied to the curved spacetime using the tetrad. 
Then the forward SDE with $dt>0$ is given by 
\begin{eqnarray}
d\hat{x}^{i}_t =  {u}^{i}_{+} (\hat{x}_t) dt + \sqrt{\frac{\hbar}{M}} \underline{e}^{i}_a (\hat{x}_t) \circ_s d\hat{W}^a_t\, ,
\end{eqnarray}
where $\hat{x}_t$ represents $(x^0_t, \hat{\vec{x}}_t)$ and the Stratonovich definition of the product is given by 
\begin{eqnarray}
f(\hat{x}_t) \circ_s d\hat{W}^a_t = f(\hat{x}_{t+ dt/2}) d\hat{W}^a_t\, ,
\end{eqnarray}
for an arbitrary smooth function $f(x)$.
It should be noted that Leibniz's rule of differential for stochastic quantities is formally held when the Stratonovich definition is applied \cite{book:gardiner}.
The standard Wiener process $\hat{W}^a_t$ has only spatial components ($\hat{W}^{0}_t = 0$) which satisfy
\begin{eqnarray}
E[d\hat{W}^a_t] &=& 0\, , \label{eqn:w1}\\
E[(d\hat{W}^a_t)(d\hat{W}^b_{t^\prime})] &=& |dt| \delta^{a b} \delta_{t,t^\prime}\ \ \ (a,b \neq 0)\,  , \label{eqn:w2}
\end{eqnarray}
where $E[~]$ represents the stochastic ensemble average.
The purpose of SVM is to find the unknown smooth function $u^i_+(x)$ 
by the optimization.

The change of the tetrad is determined by 
the Levi-Civita-Ito stochastic parallel transport \cite{ito}, 
\begin{eqnarray}
d\underline{e}^{\mu}_a (\hat{x}_t) 
=
-\Gamma^\mu_{\nu \delta} (\hat{x}_t) \underline{e}^\delta_a (\hat{x}_t) \circ_s d\hat{x}^\nu_t\, , \label{eqn:ev} 
\end{eqnarray}
where $\Gamma^\mu_{\nu \delta}(x)$ is the Christoffel symbol.
The length of the transported vector is conserved in this definition. 
Note however that a different stochastic transport is considered in 
Nelson's stochastic mechanics of the curved space \cite{zam-review, dankel-71,dohrn-77,dohrn-78} where, thus, the length is not conserved.

In the formulation of the variational principle, we should fix not only an initial condition but also a final condition. 
This implies that the forward SDE alone is not sufficient \cite{zam-review,koide-review}.  
We further introduce the backward SDE for $dt<0$ as 
\begin{eqnarray}
d\hat{x}^{i}_t =  {u}^{i}_-(\hat{x}_t) dt + \sqrt{\frac{\hbar}{M}} \underline{e}^{i}_a (\hat{x}_t) \circ_s d\underline{\hat{W}}^a_t\, ,
\end{eqnarray}
where ${u}^{i}_-(x)$ is another unknown function and $\underline{\hat{W}}^a_t$ is another Wiener process 
which satisfies the same correlations as $\hat{W}^{a}_t$.
The backward SDE should correspond to the time-reversed process of the forward SDE.  
Thus there exists a condition associating $u^i_-(x)$ with ${u}^i_+(x)$.
To find it, we derive the two Fokker-Planck equations for the probability density defined by 
\footnote{For the sake of simplicity, we omitted the initial distribution of the particle here but it does not affect our formulation.}, 
\begin{eqnarray}
\rho(x) 
&=&  \int \frac{c dt}{\sqrt{-g}} \, E[\delta^{(4)}(x^\mu - \hat{x}^\mu_t)]\, , 
\end{eqnarray}
from the forward and backward SDEs, independently. 
These Fokker-Planck equations should be equivalent and then, using Eq.\ (\ref{eqn:cond-geo}),  
we find the consistency condition, 
\begin{eqnarray}
u^i_+ (x) &=& {u}^i_- (x) + \frac{\hbar}{M} g^{ij} \partial_j \ln \rho (x)\, .  
\label{eqn:cc}
\end{eqnarray}  
See also Refs.\ \cite{zam-review,koide-review}.
Thus the two Fokker-Planck equations are shown to be reduced to the same equation of continuity \cite{zam-review,koide-review}, 
\begin{eqnarray}
\nabla_\mu (\rho(x) v^\mu (x)) = 0\, , \label{eqn:fp}
\end{eqnarray}
where  $\nabla_\mu$ represents the covariant derivative and 
\begin{eqnarray}
v^\mu (x)= (v^0 , v^{i}(x)) = \left( c, 
 \frac{u^i_+(x)  + {u}^i_-(x)}{2} \right)\, .
\end{eqnarray}

The stochastic particles follow zigzag paths 
and thus the standard definition of the particle velocity is not applicable. 
The possible time differentials are studied by Nelson \cite{nelson}: one is the mean forward derivative,
\begin{equation}
D_+  f(\hat{x}_t)  = \lim_{dt \rightarrow0+} E \left[  \frac{ f(\hat{x}_{t + dt}) -
f(\hat{x}_t)}{dt} \Big| \mathcal{P}_{t} \right]\,  ,
\end{equation}
and the other the mean backward derivative,
\begin{equation}
{D}_-  f(\hat{x}_t)  = \lim_{dt \rightarrow0-} E \left[  \frac{ f(\hat{x}_{t + dt}) -
f(\hat{x}_t)}{dt} \Big| \mathcal{F}_{t} \right]\, .
\end{equation}
These expectation values are conditional averages, where $\mathcal{P}_{t}$ ($\mathcal{F}_{t}$) indicates to fix values of $\hat{x}^{i}_{t^\prime}$ for
$t^{\prime}\le t~~(t^{\prime}\ge t)$. 
For the $\sigma$-algebra of all measurable events of $\hat{x}_t$, $\mathcal{P}_{t}$ and $\mathcal{F}_{t}$
represent an increasing and a decreasing family of sub-$\sigma$-algebras, respectively. 
These derivatives are connected through the stochastic partial integration \cite{koide-review},
\begin{eqnarray}
\int^{b}_a ds E[\hat{Y}_s D_+ \hat{X}_s] &=& - \int^{b}_a ds E[\hat{X}_s D_- \hat{Y}_s] 
\nonumber \\
&& 
+ \int^b_a ds \frac{d}{ds} E[\hat{X}_s \hat{Y}_s]\, .
\end{eqnarray}

As another new aspect in the formulation for the curved spacetime, 
$dy^{a} = dx^{\mu} \overline{e}^{a}_{\mu} (x)$ 
is generalized for the stochastic trajectories. Although it is not unique, we adapt the generalization using the Stratonovich definition,
\begin{eqnarray}
dy^{a}(\hat{x}_t) = d\hat{x}^{\mu}_t \circ_s \overline{e}^{a}_{\mu} (\hat{x}_t)\, ,
\end{eqnarray}
leading to
\begin{eqnarray}
 \underline{e}^{i}_a (\hat{x}_t) D_{\pm} y^{a}(\hat{x}_t) = u^{i}_\pm (\hat{x}_t)\, .
\end{eqnarray}
Similarly, for a smooth vector function $A^\mu(x)$, we find 
\begin{eqnarray}
\lefteqn{\underline{e}^{i}_a(\hat{x}_t) D_{\pm} (A^{\mu} (\hat{x}_t) \overline{e}^{a}_{\mu}(\hat{x}_t)) } \nonumber \\
&&\hspace*{-0.5cm}=\underline{e}^{i}_a(\hat{x}_t)  E\left[ \frac{dA^\mu(\hat{x}_t)}{dt} \circ_s \overline{e}^{a}_\mu (\hat{x}_t) + A^\mu (\hat{x}_t)\circ_s \frac{d\overline{e}^{a}_\mu(\hat{x}_t)}{dt}  \Big| \mathcal{ P}_t (\mathcal{F}_t) \right] \nonumber \\
&&\hspace*{-0.5cm}= \left( c \nabla_0
+ u^j_{\pm} (\hat{x}_t)\nabla_j \pm  \frac{\hbar}{2M} g^{jk}(\hat{x}_t) \nabla_j \nabla_k \right) 
A^i (\hat{x}_t)\, .
\end{eqnarray}
Here we used 
$\underline{e}^{\nu}_a  \circ_s d \overline{e}^{a}_\mu 
+ \overline{e}^{a}_\mu  \circ_s d \underline{e}^{\nu}_a =0$.

Let us apply these definitions to the variation of the stochastic action,  
\begin{eqnarray}
I= \int^{t_f}_{t_i}dt\, E\left[ L \right]\, .
\end{eqnarray}
As the stochastic Lagrangian, we consider a single-particle system of the mass $M$, 
\begin{eqnarray}
L 
= \frac{M}{4} \sum_{z=+,-} ( D_z \hat{y}^{a}) \eta_{ab} ( D_z \hat{y}^{b})   - V(\hat{x}_t)\, ,
\end{eqnarray}
where $\hat{y}^{a} = y^{a}(\hat{x}_t)$ and $V(x)$ is a potential energy. 
In the stochastic systems, the tetrad is considered to be more fundamental than the metric and the Lagrangian is expressed by the tetrad.
The kinetic term is replaced by the average of the contributions from the mean forward and backward derivatives. 
See Refs.\ \cite{koide-review,koide-12,koide-18} for other choices.
This reduces to the standard classical Lagrangian in the vanishing limit of $\hbar$.

The variation of the trajectory is defined by 
\begin{eqnarray}
\hat{x}^{i}  \longrightarrow \hat{x}^{i}_t + \delta f^{i}(\hat{\vec{x}}_t,t)\, , 
\end{eqnarray}
where the infinitesimal smooth function satisfies $\delta f^{i} (\vec{x},t_i) = \delta f^{i} (\vec{x},t_f) =0$. 
We find the optimized solution for any choice of 
$\delta f^{i} (\vec{x},t)$ and also for any stochastic distribution of $\hat{x}^{i}_t$ \cite{yasue-81,zam-review,koide-review}. 
Then $v^{i}(x)$ is given by the solution of the following equation, 
\begin{eqnarray}
\lefteqn{(v^0 \nabla_0 + v^j  \nabla_j) v^i + \frac{g^{ij}}{M} \partial_j V} && \nonumber \\
&& = \frac{ g^{ij} \hbar^2}{2M^2}
\left( 
 \partial_j \frac{1}{\sqrt{\rho}} \Delta_{LB} \sqrt{\rho}
- \frac{1}{2}{R_{j}}^k \partial_k \ln \rho \right)\, .  \label{eqn:vari-particle}
\end{eqnarray}
Here $\Delta_{LB} = g^{ij} \nabla_i \partial_j$ is the Laplace-Bertrami operator but the sum runs only for $i,j=1,2,3$ 
because of the correlation defined by Eq.\ (\ref{eqn:w2}).
The first term on the right hand side is known as the quantum potential which exists even in the Euclidean space and 
produces various non-trivial quantum behaviors \cite{book:holl,sanz-15}. 
The second term with the Ricci tensor 
${R_{j}}^k = g^{kl} {R_{jl}}=g^{kl}  {R^{\mu}}_{jl\mu}$ \cite{book:weinberg} is the new term induced by the interplay between quantum fluctuation and spacetime curvature. 
This is the term which we have called quantum-curvature (QC) term.

To cast Eq.\ (\ref{eqn:vari-particle}) into the form of the Schr\"{o}dinger equation, $v^i$ should be expressed in terms of the velocity potential which becomes the phase of the wave function. 
See Appendix \ref{app:1}.  
However, because of the possible $x$ dependence in the Ricci tensor, 
Eq.\ (\ref{eqn:vari-particle}) is not generally represented by the velocity potential.
Therefore we cannot introduce the wave function using the standard procedure in quantum hydrodynamics.
This seems to suggest that our quantum mechanics in the curved spacetime is not 
expressed in terms of the wave function and hence 
the representation of quantum states with a linear vector space is not possible.

It is straightforward to apply the above formulation to many-body systems when the interaction is represented by potential. 
For the sake of later convenience, however, we rather consider continuum media 
which is the coarse-grained description of the many-body systems.
The behavior of the non-relativistic simple fluid is described 
by the mass density and the velocity field.
The corresponding  stochastic Lagrangian density is obtained from the classical Lagrangian density for the ideal fluid.
In the Lagrangian coordinates, it is expressed as \cite{koide-12,koide-18,koide-15}  
\begin{eqnarray}
{\cal L}  &=&  \rho_{M0} ({\xi}) \nonumber \\
&& \hspace{-1cm} \times \left[
 \frac{1}{4} \sum_{z=+,-} (D_z \hat{y}^{a})\eta_{ab}( D_z \hat{y}^{b}) 
- \frac{\varepsilon_m [\rho_{M0}({\xi}) / J(\partial \hat{x}_t)]}{\rho_{M0}({\xi}) / J(\partial \hat{x}_t)}
\right]\, , \ \ \ \ \
\end{eqnarray}
where $\rho_{M0}({\xi})$ is the initial mass distribution, $\varepsilon_m$ is the internal energy density and 
$J(\partial \hat{x}_t) = 
 {\rm det} \left|
\partial y(\hat{x}_t) / \partial y(\xi)
\right|$.
Note that the particle trajectory in the previous discussion is replaced with that of the fluid element, 
$\hat{x}^\mu_t \rightarrow \hat{x}^\mu_t({\xi})$ where $\xi$ denotes the initial position of the fluid element.
The mass density of the simple fluid is given by $\rho_M(x) = M \rho(x)$ and then the consistency condition (\ref{eqn:cc}) 
is still held by replacing $\rho(x)$ with $\rho_M(x)$. 
Note that the Gross-Pitaevskii equation is derived from ${\cal L} $ when the Euclidean SVM is applied \cite{koide-12,koide-18}.

For the sake of later convenience, we express the optimized result with the fluid momentum density as
\begin{eqnarray}
\nabla_0 (\rho_{M} v^{i}v^{0} ) 
= - \nabla_j T^{ij}_m\, ,
\end{eqnarray}
where the fluid stress tensor is defined by 
\begin{eqnarray}
T^{ij}_m
=
\rho_M v^{i} v^j + g^{ij} P_m - \frac{\hbar^2}{4M^2} g^{i\alpha} g^{j\beta}\rho_M \nabla_\beta \partial_\alpha \ln \rho_M\, . \ \ \label{eqn:svm-rw}
\end{eqnarray}
The adiabatic pressure $P_m$ is \cite{koide-12}
\begin{eqnarray}
P_m = -\frac{d}{d\rho^{-1}_M} \frac{\varepsilon_m(\rho_M)}{\rho_M}\, ,
\end{eqnarray}
and the QC term and the quantum potential are unified to the last term. 
The suffix $m$ indicates the contribution from matter.
Note that $\rho_{M} v^{i} v^{0}$ and $T^{ij}_m$ form a part of the energy-momentum tensor.
The quantum potential is not sufficient and the QC term is necessary to express the conserved energy-momentum tensor.

\section{Effect of QC term}

As was discussed, the QC term can prohibit us from introducing the wave function and then the 
standard formulation of quantum mechanics needs revision. 
Therefore the adequacy of the introduction of such a term should be investigated carefully. 

The typical example of the curved spacetime systems appears in cosmology.
To evaluate the effect of the QC term, 
we apply the FRW metric \cite{book:weinberg}, $g_{\mu\nu} = {\rm diag} \left( -1, a^2/(1-Kr^2), a^2 r^2, a^2 r^2 \sin^2 \theta \right)$,
where $a$ is the FRW scale factor and $K$ is a parameter associated with the geometry, which takes the value $1$ (spherical), $0$ (Euclidean) 
or $-1$ (hyperspherical). 
We further consider that the distribution of matter with large scales is homogeneous and isotropic. 
Then, dropping the velocity field and the spatial derivative terms in Eq.\ (\ref{eqn:svm-rw}), 
the stress tensor becomes
\begin{eqnarray}
T^{ij}_m 
&=&
g^{ij} \left( P_m   -  \frac{c^4}{8\pi G} \Lambda_{QC}  \right), \\
 \Lambda_{QC}
&=& 
- \frac{2\pi \hbar^2 G}{M^2 c^5} H_0 \partial_0 \rho_M\, , 
\label{eqn:tiso1}
\end{eqnarray}
where $H_0 = a^{-1}(da/dt)$ is the Hubble constant and the dimension of $\Lambda_{QC}$ is the same as that of the cosmological constant $\Lambda$.
The second term in $T^{ij}_m$ gives the negative contribution to the pressure \footnote{This term contains also the contribution from the quantum potential.}. 
To see this, we use the energy conservation \cite{book:weinberg} given by   
\begin{eqnarray}
\partial_0 \rho_M \approx c^{-2} \Omega_m \partial_0 \varepsilon
= -3 c^{-3} \Omega_m H_0 (\varepsilon + P)\, , \label{eqn:con-mass}
\end{eqnarray} 
where $\varepsilon$ and $P$ are the total energy density 
of the universe and the corresponding pressure, respectively. 
The energy ratio $\Omega_m$ is defined by $\rho_M c^2 /\varepsilon$. 
We assumed that the time dependence of $\Omega_m$ is small.
Then $\Lambda_{QC}$ becomes a positive function,
\begin{eqnarray}
\Lambda_{QC} 
=  \frac{6\pi G}{c^4} \alpha \Omega_m (\varepsilon +  P)\, ,
\end{eqnarray} 
where the adimensional parameter is defined by 
\begin{eqnarray}
\alpha =  \left( \frac{\hbar H_0}{M c^2} \right)^2 \approx 10^{-84}\, , 
\end{eqnarray}
with $M$ being chosen to be, for example, the mass of the hydrogen atom $M = 1.67 \times 10^{-27}$ kg.
That is, $\Lambda_{QC}$ induces the effect analogous to $\Lambda$ which contributes
to the accelerating expansion of the universe.

Whereas the above qualitative behavior is consistent with the observation, 
the magnitude is far from the observed value. 
Substituting the critical density to $\varepsilon + P$, the order is estimated by  
$\Lambda_{QC} \approx 10^{-137}$ m$^{-2}$ with $\Omega_m = 0.1$, 
which is too small compared to the accepted value, $\Lambda \sim 10^{-52}$ m$^{-2}$ \cite{de-review1,de-review2,stocker-18}. 
The situation is not much improved even if we consider a lightweight particle of the dark matter candidate like axion.
It should be however noted that $\Lambda_{QC}$ in Eq.\ (\ref{eqn:tiso1}) can be inhomogeneous 
and the above estimate is not appropriate for the quantitative discussion of the accelerating expansion. 
See also the later discussion.

On the other hand, we observe that, applying Eq.\ (\ref{eqn:tiso1}) to a localized mass distribution, 
the second term in $T^{ij}_m$ 
gives a negative contribution for $\partial_0 \rho_M <0$ and a positive constitution for $\partial_0 \rho_M >0$, respectively. 
Therefore the QC term suppresses the diffusion of the mass distribution and then the gravitational force seems to be effectively enhanced. 
Such a behavior is analogous to a part of the effects expected by the dark matter \cite{feng,zurek}.

It is known that the expectation value of the energy-momentum tensor operator can be calculated in the standard quantum field theory, 
leading to the correction to the cosmological constant induced by the interplay between 
quantum fluctuation and spacetime curvature (see, for example, \cite{fischetti79}). 
However the QC term is not reproduced in such an approach because it is not expressed in the form of the expectation value of an operator.

\section{Concluding remarks}

\begin{table}[t]
\caption{SVM quantizations in different spacetimes}
\label{table:1}
\begin{tabular}{|c||c|c| }
\hline
 Spacetime & Schr\"{o}dinger eq. & Quantum hydro. \\
 \hline
\hline 
 Euclidean   & YES   & YES \\
 Curved  &   NO & YES  \\
 \hline
\end{tabular}
\end{table}

In this paper, 
the stochastic variational method was generalized to the non-relativistic curved spacetime systems for the first time.
Then we can discuss quantization without assuming the existence of the wave function a priori.
This is applicable not only to non-relativistic particle systems but also to continuum media.
We further showed that the interplay between quantum fluctuation and spacetime curvature induces the quantum-curvature term 
which can prevent us from deriving the Schr\"{o}dinger equation by introducing the wave function, 
but is necessary to find the conserved energy-momentum tensor. 
As an example of the curved spacetime systems, the Friedmann-Robertson-Walker metric was applied. 
The QC term is consistent in the sense that it does not violate the standard scenario of cosmology: 
the effect of the QC term contributes to the cosmological acceleration 
although its quantitative influence is too small in the present non-relativistic toy model.

It should be noted that the application to the cosmological problem is not our principal purpose. 
We intended to find a possibility for the large modification of quantum mechanics.
In Table \ref{table:1}, the SVM quantizations in the Euclidean and curved spacetime systems are summarized. 
As is well-known, 
quantum mechanics in the Euclidean spacetime is described by quantum hydrodynamics 
which can be reexpressed in the form of the Schr\"{o}dinger equation \cite{yasue-81, zam-review,koide-review,book:holl}.  
On the other hand, when we apply SVM to curved spacetime systems, 
the obtained quantum hydrodynamics is not necessarily cast into the form of the Schr\"{o}dinger equation.
In this discussion, we assumed that the standard procedure to obtain the Schr\"{o}dinger equation 
from quantum hydrodynamics is applied.
There may exist another way to derive the Schr\"{o}dinger equation for any curved spacetime, 
but such a method is not yet found. See Appendix \ref{app:1}.
Moreover, the background spacetime geometry was assumed 
to be given by an external field so far.
Thus 
the wave function may re-appear in a larger function space where the fluctuation of the geometry is 
taken into account.

There are attempts to gain a new insight for the interplay between quantum fluctuation 
and curved geometry in the Lab. 
For example, the Bose-Einstein condensate is regarded as an analogue black hole \cite{unruh,lahav,stein}. 
The quantum superposition of spacetime geometries 
will be observed through the gravitational entanglement of mesonic particles which is called Bose-Marletto-Vedral (BMV) 
effect \cite{bose, mar,chri}.  
Our toy model may be applied to study these kinds of phenomena.

The existence of the propagations from the past and the future is a remarkable feature of SVM. 
Such a property can be however found even in different branches of physics. 
For example, the interaction between charged particles is described 
by the average of retarded and advanced fields in Wheeler-Feynman's absorber theory \cite{wf}. 
Moreover, there is a proposal to interpret quantum mechanics in a time-symmetric manner \cite{abl,ahr-book,elizer}.
Interestingly, the similar idea was already introduced in the probability theory by Schr\"{o}dinger in 1931 \cite{sch31,sch32}.
This is known as the reciprocal process or Schr\"{o}dinger bridge problem \cite{ber32,jam74,pavan}.
The relation between SVM and the reciprocal process is discussed in Refs.\ \cite{eqm1,eqm2}.
The reciprocal process is further useful to understand the nonequilibrium work relation in statistical physics \cite{koide-rp}.

There exist various effects which have not yet been discussed.
For example, 
because of the single-valuedness of wave function, quantum hydrodynamics is considered to require 
an additional condition analogous to the Bohr-Sommerfeld quantization condition. See Refs.\ \cite{tw1,tw2} and Sec. 3.2.2 in Ref.\ \cite{book:holl}.
The corresponding condition for the present model is not yet investigated.
The relativistic effect may modify the QC term and then the Wiener process 
in the present work may be substituted by the Poisson process \cite{kudo}. 
To have the consistent back reaction between the quantized matter and 
the classical curved spacetime, the quantum-classical hybrids should be considered \cite{qch-koide,elze}. 
It is also interesting to study the vacuum energy and the particle creation in the field theory \cite{svm-field,ramos1,ramos2,hu}. 
These are challenges for the future.

Finally we would like to emphasize again that the Hilbert space is not necessarily 
a priori requirement in quantization and 
thus it is important to pursue the possible modifications of quantum mechanics from a wide perspective 
not only in curved spacetime systems but also in open quantum systems 
\cite{koide-18,koide-fp,koide-rp,jpg2,jpg,mostafa-10}.

\vspace{1cm}

The authors thank V.\ Brasil, M.\ O.\ Calv\~{a}o, H.-T.\ Elze, J.\ Schaffner-Bielich, A.\ K.\ Kohara and R.\ O.\ Ramos for useful discussions and comments, 
and acknowledge the financial support by CNPq (307516/2015-6,303468/2018-1), FAPERJ, CAPES and PRONEX. 
A part of the work was developed under the project INCT-FNA Proc.\ No.\ 464898/2014-5.

\appendix

\section{Phase of wave function} \label{app:1}

There is a spatial geometry where we can still introduce the Schr\"{o}dinger equation in curved geometries.
Let us consider the Ricci tensor given by the diagonal form,
\begin{eqnarray}
{R_j}^k =- 2 \gamma \delta^k_j\, , \label{eqn:hom-ricci}
\end{eqnarray}
where $\gamma$ is a constant.
This is observed, for example, in the geometry of hyperspherical surface.

By introducing a scalar field $\theta$ by 
\begin{eqnarray}
m v^i = \hbar g^{ij}\partial_j \theta\, , \label{eqn:phase}
\end{eqnarray}
Eq. (\ref{eqn:vari-particle}) is reexpressed as the equation for $\theta$, 
\begin{eqnarray}
\lefteqn{\hbar \partial_t \theta + \frac{\hbar^2}{2m} (\partial_i \theta) g^{ij} (\partial_j \theta) + V } \nonumber \\
&& =
\frac{\hbar^2}{2m} \frac{1}{\sqrt{\rho}}\Delta_{LB} \sqrt{\rho} + \frac{\hbar^2}{2m} \gamma \ln \rho \, .
\end{eqnarray}
We further introduce the wave function $\phi$ using $\theta$ as the phase, 
\begin{eqnarray}
\phi = \sqrt{\rho} e^{i\theta}\, .
\end{eqnarray}
Then $\phi$ is described by the following non-linear Schr\"{o}dinger equation,
\begin{eqnarray}
i\hbar \partial_t \phi = \left[
-\frac{\hbar^2}{2m}\Delta_{LB} + V - \frac{\hbar^2}{2m}\gamma \ln |\phi|^2
\right]\phi\, .
\end{eqnarray}

By analyzing the above derivation, one notices that, to obtain the equation for $\theta$, the QC term is expressed as \begin{eqnarray}
g^{ij} {R_j}^k \partial_k \ln \rho = g^{ij} \nabla_k ( {R_j}^k \ln \rho)\, .
\end{eqnarray}
However, because of the possible $x$ dependence in ${R_j}^k $, this is not generally satisfied.

Is it possible to extend the above argument to find the Schr\"{o}dinger equation for general ${R_j}^k$?
For example, we replace Eq.\ (\ref{eqn:phase}) with
\begin{eqnarray}
m v^i = \hbar g^{ij}\partial_j \theta + B^{i}\, , 
\end{eqnarray}
where $B^i$ is a vector field.
Substituting this into Eq.\ (\ref{eqn:vari-particle}) and choosing the form of $B^i$ appropriately, 
it might be possible to find the equation of $\theta$ for any ${R_j}^k$. 
However, then, the momentum, which is given by the gradient of $\theta$ in quantum hydrodynamics, 
does not coincide with the inertia defined by the product of mass and velocity. 
This occurs for charged-particles interacting with gauge fields. 
See Eq.\ (58) of Ref.\ \cite{koide-12}. 
In the present case, the momentum is equivalent to the inertia at least in the classical limit and thus 
it is not easy to justify such a modification.
The answer to the above question is negative at the present moment.

\end{document}